\documentclass[showkeys,showpacs,twocolumn]{revtex4}
\usepackage{amsmath}
\usepackage{amscd}
\usepackage{graphicx}
\usepackage{subfigure}
\usepackage{appendix}
\usepackage{amsfonts}
\usepackage{multirow}
\usepackage{bm}
\usepackage{float}

\begin{document}

\title[Short Title]{Accelerated and noise-resistant generation of a high-fidelity steady-state entanglement with Rydberg atoms}

\author{Ye-Hong Chen$^{1,2}$}
\author{Zhi-Cheng Shi$^{1,2}$}
\author{Jie Song$^{3}$}
\author{Yan Xia$^{1,2,}$\footnote{E-mail: xia-208@163.com}}
\author{Shi-Biao Zheng$^{1,2}$}

\affiliation{$^{1}$Department of Physics, Fuzhou University, Fuzhou 350116, China\\
             $^{2}$Fujian Key Laboratory of Quantum Information and Quantum Optics (Fuzhou University), Fuzhou 350116, China\\
             $^{3}$Department of Physics, Harbin Institute of Technology, Harbin 150001, China}

%\tableofcontents

\begin{abstract}
  {Based on Lyapunov control}, a scheme is proposed to accelerate the dissipation dynamics for the generation of high-fidelity entanglement between two Rydberg atoms
  in the context of cavity quantum electrodynamics (QED). We first use the quantum Zeno dynamics and Rydberg antiblockade
  to find a unique steady state (two-atom singlet state) for the system. Then, applying additional coherent control (ACC) fields
  to improve the evolution speed of the dissipative system.
  The ACC fields are designed based on the target state and they vanish gradually along with
  increasing of the fidelity thus the system is guaranteed to be finally stable.
  Besides, the current accelerated scheme is checked to be robust against systematic and amplitude-noise errors.
\end{abstract}


\maketitle
\section{Introduction}
There is now growing interest in obtaining accelerated dynamics because fast and noise-resistant schemes
are natural requirements in quantum information processing.
The accelerated dynamics is also expected to have the ability to restrain the accumulated negative effect caused by dissipation during a long time evolution.
For instance, an approach named ``Shortcuts to adiabaticity'' (STA) \cite{ETSISMGMMACDGOARXCJGMAmop13} combining
the advantages of (fast) resonant pulses and (robust) adiabatic techniques has attracted many attentions in recent years \cite{Prl105123003,Prl109100403,Prl111100502,Prl116230503,Pra93052109,Pra84031606Epl9660005,arXiv160105551,Pra86033405,Pra84023415,Pra83062116,Pra89033403,
Pra93012311,Pra87043402,Njp16015025,Pra90060301}
and been applied in fields including fast population transfer \cite{Pra89033856,Nc712479,Np13330}, fast entanglement generation \cite{Pra94052311,Pra96022314,Pra89012326},
fast quantum computation \cite{Pra91012325}, and so on \cite{Pra8705250289063412,Njp16053017,Pra94043623,Njp18012001,Pra95022332}.
However, ``shortening the time always implies an energy cost'' \cite{Prl118100602,Prl118100601,Pra86033405,Pra93052109,Pra93012311},
%one can find the populations for the intermediate states based on STA are generally larger as compared to those based on adiabatic techniques.
one can usually find the intermediate states are populated into a relatively high level by using STA for the goal of accelerating \cite{Prl109100403,Prl116230503,Pra93052109}.
In recent schemes for fast entanglement generation based on STA in atomic systems \cite{Pra94052311,Pra96022314,Pra89012326},
since the intermediate states are excited, the negative effect caused by
dissipation does not decrease remarkably even though the evolution time is significantly shortened.
There exists a trade-off between the total evolution time and the populations of excited states \cite{Prl118100601,Prl118100602}.
%On solving the problem caused by dissipation,
That is, directly shortening the evolution time seems unable to restrain the negative
effect caused by dissipation in atomic systems for quantum entanglement generation.

On the other hand, rather than considering dissipation as a
detrimental effect, recent studies have changed the view for
dissipation due to the fact that the environment can be
used as a resource for quantum computation and entanglement
generation \cite{Prl106090502,arXiv10052114,Njp11083008,Nat4531008,Jpa41065201,Np5633,Prl107120502}. Currently, there are several
representative schemes creating steady entanglement
of high quality by dissipation \cite{Prl110120402,Prl111033606,Pra96012328,Pra95062339,
Nat504419,Prl117040501,arXiv11101024,Pra84022316,Pra83042329,Pra82054103,Prl89277901,
Prl117210502,Pra84064302,Nat504415,Prl111033607,Pra95022317,Npto10303,Prl115200502,Prl107080503}. For instance,
two groups independently proposed theoretical schemes to prepare high-fidelity steady-state
entanglement between a pair of Rydberg atoms with dissipative
Rydberg pumping \cite{Prl111033606,Prl111033607}.
In 2011, Krauter \emph{et al.}
experimentally realized a steady-state entanglement of two macroscopic objects by dissipation \cite{Prl107080503}.
In general, by using dissipation dynamics to generate atomic entanglement in cavity QED systems,
the fidelity $F$ of the target state is in a relationship $(1-F)\propto C^{-1}$ with the cooperativity $C=g^{2}/(\gamma\kappa)$ \cite{Prl106090502},
where $g$ is the atom-cavity coupling strength, $\gamma$ is the atomic decay rate, and $\kappa$ is the cavity decay rate.
A large cooperativity is always necessary in order to obtain a high-fidelity entanglement.
However, a large cooperativity leads to a very long convergence time (total evolution time)
that is also unwelcome \cite{Prl106090502,Pra85032111,Pra84064302}. It would be a serious issue to realize large-scale integrated computation if
taking too long for entanglement generation.
We are thus guided to ask, is it possible to accelerate the slow dissipation dynamics without losing its advantages?

The idea of combining advantages of resonant pulses and adiabatic techniques in STA inspires us that
combining advantages of dissipation dynamics and another (fast) dynamics maybe a good idea to solve the problem.
Therefore, in this paper, we combine dissipation dynamics with coherent unitary dynamics and propose a promising
scheme for an accelerated and dissipation-based entanglement generation.
We add target-state-related additional coherent control (ACC) fields into the
dissipation process. The intensities of the ACC fields are designed to decrease with the increasing of fidelity for the target state.
{To realize such an idea, we use Lyapunov control which may
have the ability to shorten the convergence time of an open system as pointed out by Yi \emph{et al.} in Ref. \cite{Pra80052316}.
Lyapunov control is a form of local optimal control
with numerous variants \cite{Ddbook,Pra80052316,A4498,Njp11105034,Pra91032301,Pra88063823} and has been
used to manipulate open quantum systems \cite{Pra80052316,Pra80042305,Pra86022321,OI231650005}.}
In this case, the evolution of the system can be understood as two stages:

(i) The first stage is mainly governed by the ACC fields. The evolution in this stage is nearly unitary so that
the system can be rapidly driven to the target state with fidelity about $90\%$. In this stage,
the target state is not a steady state of the system.

(ii) The second stage is mainly governed by the dissipation dynamics. When the fidelity for the target state is $\sim 90\%$,
the intensities of the ACC fields become very small and their effects on the dynamics can be ignored.
The dissipation dynamics thus governs the system to converge to the target state with fidelity
increasing from $\sim90\%$ to $\sim100\%$. In this stage, the target state is the unique steady state of system.

Since the evolution is accelerated in the first stage, the total evolution time required in the current scheme is much shorter than that in a general
dissipation-based scheme.
This idea is verified by an atom-cavity system via quantum Zeno dynamics \cite{Prl89080401Jpa41493001} and the Rydberg antiblockade in this paper.
Regarding two typical dissipation sources in a cavity QED system, we make use of atomic decay but avoid the
effect of cavity decay based on quantum Zeno dynamics. The Rydberg antiblockade as shown theoretically in Refs. \cite{Pra95022317,Pra95062339,Pra96012328} can
accelerate the convergence rate of stationary entanglement, since the strength of antiblockade interaction is much
larger than the Rabi frequency of microwave field. The ACC fields are chosen as the easily realized classical drivings.
Their intensities are designed as functions of system's evolution speed $v$ (time derivative of fidelity).
For $t\rightarrow t_{f}$ ($t_{f}$ is the final time), the system gradually becomes stable, i.e., $v|_{t\rightarrow t_{f}}\rightarrow0$,
that guarantees the ACC fields vanish gradually along with the increasing of time.
Thence a fidelity $\sim95\%$ of steady-state entanglement is available even with evolution
time $t_{f}=250/g$.

The paper is organized as follows. In Sec. II, we guarantee a unique entangled-steady state
is existent by using quantum Zeno dynamics and Rydberg antiblockade.
In Sec. III, we define the evolution speed for the system and show how to accelerate the dissipation dynamics.
In Sec. IV, we give the analysis and discussion on the accelerated dynamics.
In Sec. V, we verify the robustness of the scheme against stochastic parameter fluctuations that generally exist in the driving fields.
Conclusion is given in Sec. VI.

\begin{figure}[t]
 \scalebox{0.16}{\includegraphics {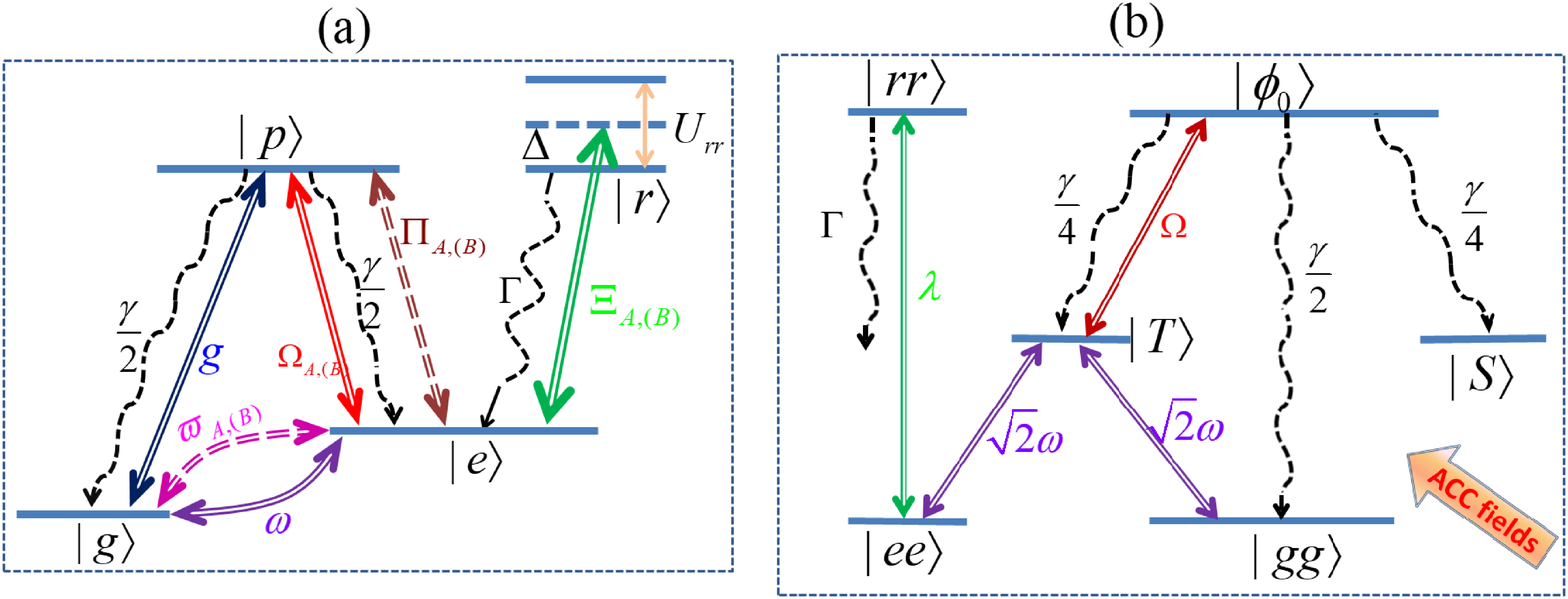}}
 \caption{
         (a) Schematic view of atomic-level configuration. The atomic transition $|g\rangle_{A,(B)}\leftrightarrow|p\rangle_{A,(B)}$
             is coupled to a quantized cavity field with coupling
             strength $g$ and the transition $|e\rangle_{A,(B)}\leftrightarrow|p\rangle_{A,(B)}$ is driven
             by two optical pumping laser with Rabi frequency $\Omega_{A,(B)}$ and $\Pi_{A,(B)}$. In addition, two microwave field of
             Rabi frequency $\omega_{A,(B)}$ and $\varpi_{A,(B)}$ are introduced to cause transition between
             ground states $|g\rangle_{A,(B)}$ and $|e\rangle_{A,(B)}$, and an extra pumping laser field
             with Rabi frequency $\Xi_{A,(B)}$ drives the atom to the high-lying
             excited Rydberg state $|r\rangle_{A,(B)}$ from state $|e\rangle_{A,(B)}$ by detuning $-\Delta$.
             $\Pi_{A,(B)}$ and $\varpi_{A,(B)}$ are the Rabi frequencies for ACC fields given according to the Lyapunov control.
         (b) The effective transitions for two-atom system. The whole system works well in the so-called Zeno $Z_{0}$
             subspace of zero occupation for the cavity mode due to the quantum Zeno dynamics.
             With the effective driving fields and decays, ultimately, the system will be stabilized into the
             state $|S\rangle$. The ACC fields mainly accelerate the transitions
             $|gg\rangle\rightarrow|T\rangle\rightarrow|\phi_{0}\rangle\rightarrow|S\rangle$ and $|gg\rangle\rightarrow|S\rangle$ to shorten the evolution time.
         }
 \label{fig0}
\end{figure}
\section{steady ground-state entanglement of two atoms}
We consider a system consisting of two $N$-type four
level Rydberg atoms (marked as atom $A$ and atom $B$), and the relevant configuration of atomic
level is illustrated in Fig. \ref{fig0} (a). We first consider that $\Pi_{A,(B)}=\varpi_{A,(B)}=0$,
the system is thus the same as that in Ref. \cite{Pra95022317}.
In the regime of Rydberg antiblockade: $U_{rr}\sim 2\Delta\gg\Xi_{A,(B)}$, the Hamiltonian \cite{Pra95022317} for the current system reads
\begin{align}\label{eq0-1}
  H_{0}=&H_{r}+H_{ac}, \cr
  H_{ac}=&\sum_{n=A,B}g_{n}|p\rangle_{n}\langle g|a+H.c., \cr
  H_{r}\approx&\sum_{n=A,B}(\Omega_{n}|e\rangle_{n}\langle p|+\omega_{n}|g\rangle_{n}\langle e|)\cr
      &+\lambda|ee\rangle\langle rr|+H.c.,
\end{align}
where $U_{rr}$ is the Rydberg-mediated interaction \cite{Prl852208,Np5110,Prl100170504,Prl102170502,Prl104010503,Rmp822313}
and $\lambda=2\Xi^{2}/\Delta$ ($\Xi_{A}=\Xi_{B}=\Xi$) is given according to the second-order perturbation theory \cite{Cjp85625}.
The dynamics of the system in this case is modeled
by Lindblad-Markovian master equation \cite{Cmp48119} as
\begin{align}\label{eq0-2}
  \dot{\rho}=&-i[H_{0},\rho]+\mathcal{L}\rho \cr
  \mathcal{L}\rho=&\sum_{k}L_{k}\rho L_{k}^{\dag}-\frac{1}{2}(L_{k}^{\dag}L_{k}\rho+\rho L_{k}^{\dag}L_{k}),
\end{align}
where the overdot means time derivative and the Lindblad operators are
\begin{align}\label{eq0-3}
  L_{n,1}&=\sqrt{\gamma/2}|g\rangle_{n}\langle p|, \
  L_{n,2}=\sqrt{\gamma/2}|e\rangle_{n}\langle p|, \cr
  L_{n,3}&=\sqrt{\Gamma}|e\rangle_{n}\langle r|,  \
  L_{4}=\sqrt{\kappa}a.\ \ \ (n=A,B)
\end{align}
The $L_{4}$ denotes the cavity decay with decay rate $\kappa$.

Then, similar as Ref. \cite{Pra95022317}, by applying the quantum Zeno dynamics (see Appendix for details)
under the strong coupling limit $\Omega_{A,(B)},\omega\ll g$,
the effective Hamiltonian takes the following concise form \cite{Pra95022317}
\begin{align}\label{eq0-9}
  H_{eff}\simeq&\Omega|T\rangle\langle\phi_{0}|+\sqrt{2}\omega|T\rangle(\langle gg|+\langle ee|)\otimes|0\rangle_{c}\langle 0|\cr
               &+\lambda|ee\rangle\langle rr|\otimes|0\rangle_{c}\langle0|,
\end{align}
where $\Omega=\Omega_{B}=-\Omega_{A}$, $|\phi_{0}\rangle=(|pg\rangle-|gp\rangle)\otimes|0\rangle_{c}/\sqrt{2}$, and $|T\rangle=(|eg\rangle+|ge\rangle)\otimes|0\rangle_{c}/\sqrt{2}$.
For the sake of simplification, we choose
$g_{A}=g_{B}=g$ and $\omega_{A}=\omega_{B}=\omega$ in obtaining the effective Hamiltonian.
The corresponding effective Lindblad operators in the Zeno $Z_{0}$ subspace are \cite{Pra95022317}
\begin{align}\label{eq0-10}
  L_{1}^{e}&=\sqrt{\frac{\gamma}{4}}|S\rangle\langle\phi_{0}|, \ L_{2}^{e}=\sqrt{\frac{\gamma}{4}}|T\rangle\langle\phi_{0}|,\cr
  L_{3}^{e}&=\sqrt{\frac{\gamma}{2}}|gg\rangle\otimes|0\rangle_{c}\langle \phi_{0}|,
\end{align}
where $|S\rangle=(|eg\rangle-|ge\rangle)\otimes|0\rangle_{c}/\sqrt{2}$,
Here the spontaneous emission of the Rydberg state is neglected according to realistic situation that $\Gamma\ll\gamma$.
Clearly from Eqs. (\ref{eq0-9}) and (\ref{eq0-10}), we find a steady state $|S\rangle$ for the effective system on account
of $H_{eff}(L_{k}^{e})|S\rangle=0$ and $(L_{k}^{e})^{\dag}|S\rangle\neq 0$ ($k=1,2,3$). Thus, for an arbitrary initial state,
it will be finally converged into the steady state $|S\rangle$ by the
process of pumping and decaying as shown in Fig. \ref{fig0} (b).
\begin{widetext}

\begin{figure}
 \scalebox{0.33}{\includegraphics {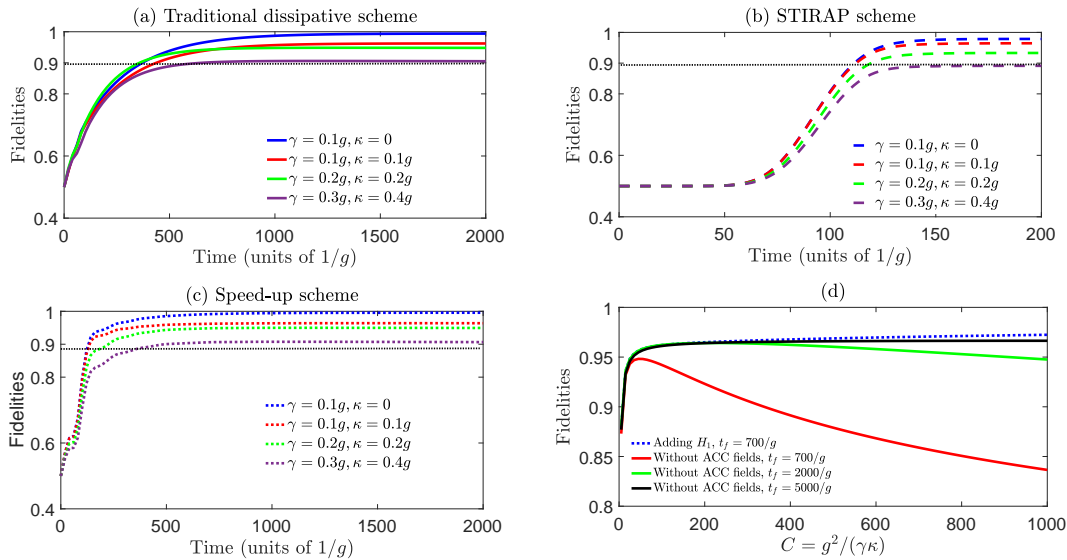}}
 \caption{
         Time evolutions of different schemes when decays are considered and initial state is $|eg\rangle\otimes|0\rangle_{c}$.
         (a) Based on a dissipation-based scheme with Hamiltonian in Eq. (\ref{eq0-1}). Parameters are $\Omega=0.07g$, $\omega=0.02g$, $\Xi=5g$, and $\Delta=100g$.
         (b) Based on a SIIRAP scheme with Hamiltonian in Eq. (\ref{eq1-1a}). Parameters are chosen as $t_{o}=20/g$, $t_{c}=35/g$, and $\Omega_{0}^{adi}=0.15g$.
         (c) Based on a speed-up scheme by adding ACC fields with $\mu_{1}=0.3g$ into the dissipative system. Parameters are the same as Fig. \ref{figcp} (a).
         (d) Comparison between dissipation-based schemes with and without ACC fields. The solid curves represent the fidelities versus $C$ of a traditional dissipation-based scheme
         with parameters the same as Fig. R2 (a). The dashed curve represent the fidelity versus $C$ when ACC Hamiltonian $H_{1}$ is applied into the system with intensity-dependent coefficient $\mu_{1}=0.3g$.
         We assume $\gamma=\kappa$ in plotting the Fig. \ref{figcp} (d).
         }
 \label{figcp}
\end{figure}
\end{widetext}

\section{The evolution speed and the principle of acceleration}
\begin{figure}
 \scalebox{0.28}{\includegraphics {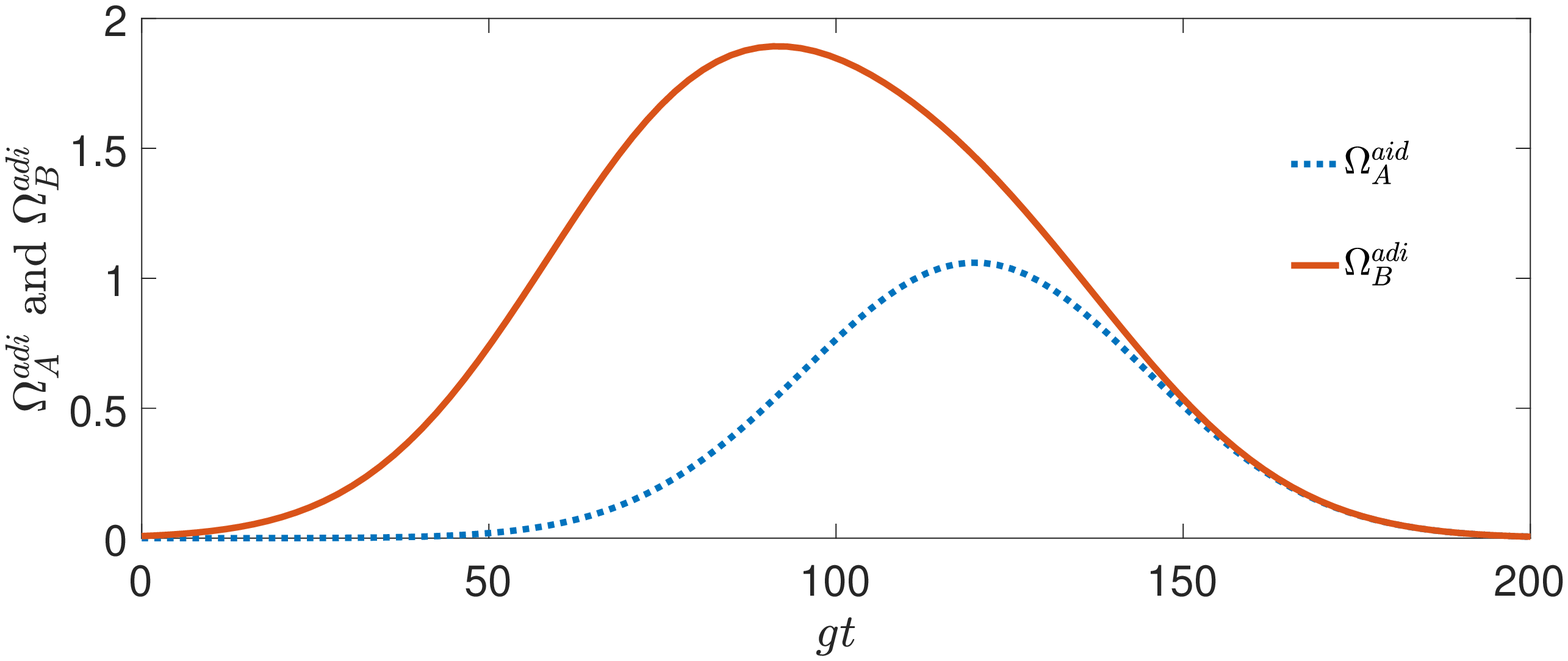}}
 \caption{
         Rabi frequencies in Eq. (\ref{eq1-1b}) of the STIRAP scheme. Parameters are chosen as $t_{o}=20/g$, $t_{c}=35/g$, and $\Omega_{0}^{adi}=0.15g$.
         }
 \label{figadi}
\end{figure}

The last section presents a method to generate an entangled-steady state $|S\rangle$ by dissipation.
However, the generation process is usually unsatisfactory slow.
We define the fidelity for the target state $|S\rangle$ as
$F=\langle S|\rho|S\rangle$. The instantaneous speed of the evolution can be thus defined as
\begin{align}\label{eq1-1}
  v=\partial_{t}F=\langle S|\dot{\rho}|S\rangle
                 =\frac{\gamma}{4}\langle\phi_{0}|\rho|\phi_{0}\rangle,
\end{align}
which depends on the spontaneous emission rate $\gamma$ and
the instantaneous population for the effective excited state $|\phi_{0}\rangle$.
The spontaneous emission rate and the population for $|\phi_{0}\rangle$ are, however, both small in the dissipation system when a high fidelity is required \cite{Prl106090502}.
As we know, the fidelity of a dissipation-based scheme is usually proportional to
the cooperativity $C$ according to the relationship $1-F\propto C^{-1}$ \cite{Prl106090502}.
The cooperativity $C$, however, is inversely proportional to decay rates.
Hence, in order to obtain a high-fidelity entanglement generation, small decay rates $\gamma$ and $\kappa$ are necessary
for a dissipation-based scheme, which lead to a long convergence time \cite{Pra95022317} [See Fig. \ref{figcp} (a)].
Fig. \ref{figcp} (a) shows the fidelity versus time with different decay rates. Obviously from the figure, the time required to
stabilize the system into the target state $|S\rangle$ increases with the decreasing of the cooperativity $C$. For example,
for $C=100$ corresponding to $\gamma=\kappa=0.1g$, the convergence time is about $t_{f}=1100/g$, while for $C=8.33$ corresponding to $\gamma=0.3g$ and $\kappa=0.4g$,
the convergence time is about $t_{f}=800/g$. However, the evolution is still slow in comparison with a STIRAP
(STIRAP is short for stimulated Raman adiabatic passage) scheme as shown in Fig. \ref{figcp} (b) which is displayed
based on an interaction Hamiltonian
\begin{align}\label{eq1-1a}
  H^{adi}=\sum_{n=A,B}\Omega_{n}^{adi}(t)|P\rangle_{n}\langle e|+g_{n}|P\rangle_{n}\langle g|+H.c.,
\end{align}
describing a system with two neutral $\Lambda$-type atoms trapped in a cavity. The time-dependent Rabi frequencies are (see Fig. \ref{figadi})
\begin{align}\label{eq1-1b}
  \Omega_{A}^{adi}=&\frac{1}{\sqrt{2}}\Omega_{0}^{adi}\exp{[-(t-t_{o}-t_{f}/2)^{2}/t_{c}^{2}]}, \cr
  \Omega_{B}^{adi}=&\frac{1}{\sqrt{2}}\Omega_{0}^{adi}\exp{[-(t-t_{o}-t_{f}/2)^{2}/t_{c}^{2}]}\cr
                   &+\Omega_{0}^{adi}\exp[-(t+t_{o}-t_{f}/2)^{2}/t_{c}^{2}].
\end{align}
According to the result of comparison between Figs. \ref{figcp} (a) and (b), it is hard to say a dissipation-based scheme is better than
a STIRAP one (even with a relatively small cooperativity $C$). When $C=8.33$, the fidelity of a STIRAP scheme is about $89\%$ which is only a little lower than
that $90\%$ of a dissipation-based scheme. However, the time required in a STIRAP scheme, i.e., $200/g$, is much shorter than that about $700/g$ in a dissipation-based scheme.
Therefore, to make sense of a dissipation-based scheme in practice, it is of significance to shorten the time required to stabilize a dissipative system.

\begin{figure}[b]
 \scalebox{0.3}{\includegraphics {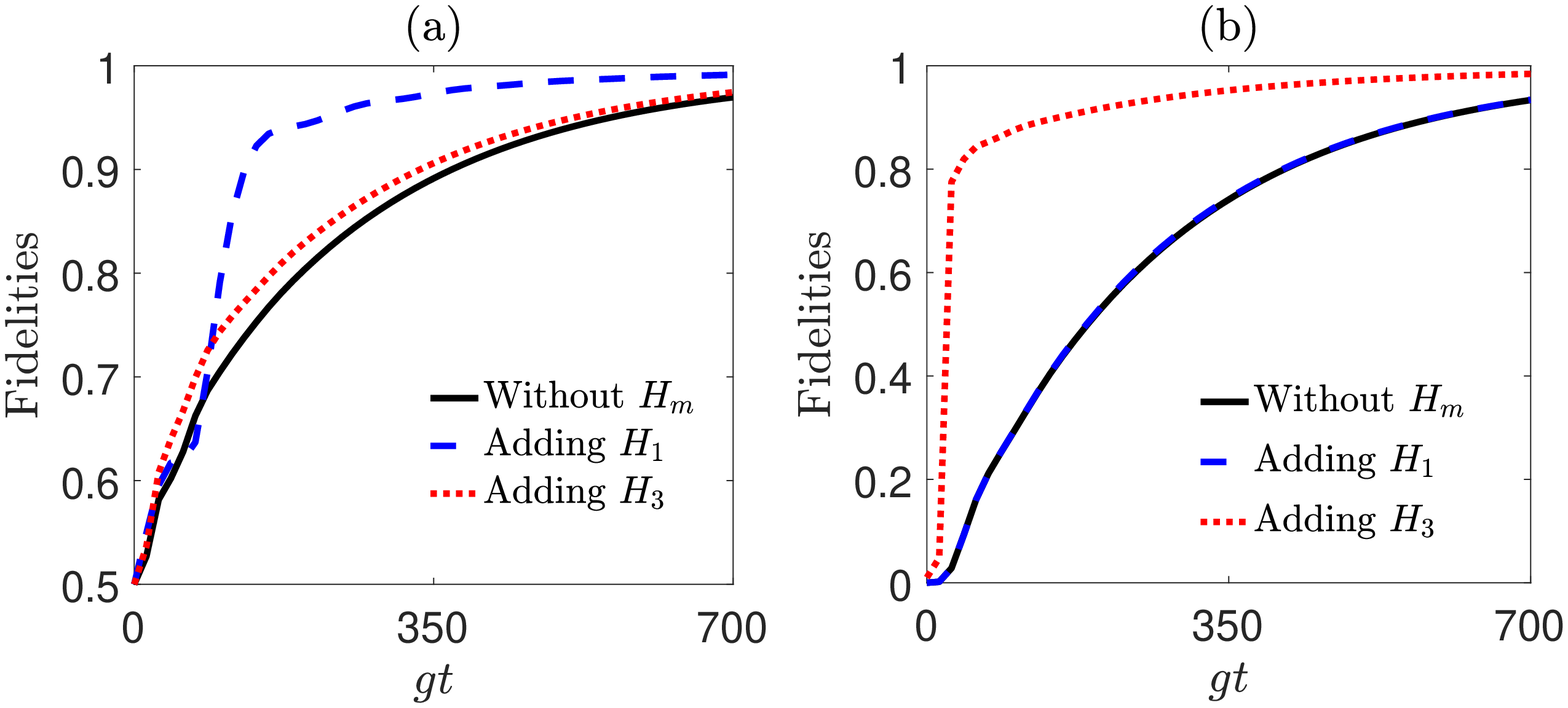}}
 \caption{
         The fidelities of singlet state $|S\rangle$ with and without adding ACC fields.
         (a) The initial state [see Eq. (\ref{eqrho0})] is chosen with $o=1$ and intensity-dependent coefficients are $\mu_{1}=0.3g$.
         (b) The initial state [see Eq. (\ref{eqrho0})] is chosen with $o=0.02$ and intensity-dependent coefficients are $\mu_{3}=0.2g$.
         Parameters are $\Omega=0.07g$, $\omega=0.02g$, $\Xi=5g$, and $\Delta=100g$.
         The decay rates are $\gamma=0.1g$, $\kappa=0$, and $\Gamma=0.001g$.
         }
 \label{fig2}
\end{figure}

We know that the fastest way to drive a quantum system to the target state is using coherent unitary dynamics.
Therefore, to accelerate the slow dissipation process for entanglement generation, we
add some ACC fields to the system.
The ACC fields should be easily realized in practice.
For the current system, the ACC Hamiltonians can be chosen as
\begin{align}\label{eq1-2}
  H_{1}=\mu_{1}|e\rangle_{A}\langle p|+H.c., \cr
  H_{2}=\mu_{2}|e\rangle_{B}\langle p|+H.c., \cr
  H_{3}=\mu_{3}|g\rangle_{A}\langle e|+H.c., \cr
  H_{4}=\mu_{4}|g\rangle_{B}\langle e|+H.c.,
\end{align}
where $\mu_{m}$ ($m=1,2,3,4$) are usually time-independent coefficients used to control the
intensities of the ACC fields. The dynamics of the effective system after adding the ACC fields is governed by
\begin{align}\label{eq1-3}
  \dot{\rho}=&-i[H_{eff}+H_{a},\rho]+\mathcal{L}\rho \cr
  \mathcal{L}\rho=&\sum_{k}L_{k}^{e}\rho (L_{k}^{e})^{\dag}-\frac{1}{2}[(L_{k}^{e})^{\dag}L_{k}^{e}\rho+\rho (L_{k}^{e})^{\dag}L_{k}^{e}],
\end{align}
with $H_{a}=\sum_{m}f_{m}(t)H_{m}$. Here, the control functions $f_{m}(t)$ can be regarded as the Rabi frequencies
for the ACC fields. In this case, the instantaneous speed of the system becomes
\begin{align}\label{eq1-4}
  v_{a}=&\langle S|\dot{\rho}|S\rangle \cr
                 =&\frac{\gamma}{4}\langle\phi_{0}|\rho|\phi_{0}\rangle-i\langle S|[H_{a},\rho]|S\rangle \cr
                 =&\frac{\gamma}{4}\langle\phi_{0}|\rho|\phi_{0}\rangle-i\sum_{m}[f_{m}(t)\langle S|[H_{m},\rho]|S\rangle].
\end{align}
Obviously, in order to improve the evolution speed, the second term in the last line of Eq. (\ref{eq1-4}) should
be ensured positive. {For this goal, according to Lyapunov control \cite{A4498}}, the control functions can be chosen as
\begin{align}\label{eq1-5}
  f_{m}(t)=-i\langle S|[H_{m},\rho]|S\rangle,
\end{align}
which are target-state-dependent functions. Beware that $\langle S|[H_{m},\rho]|S\rangle$
are purely imaginary numbers,
there is a negative sign in Eq. (\ref{eq1-5}).
The control functions mainly dependent on the definition of fidelity
for the target state, when the definition is changed, the control functions will be accordingly changed.
For example, when the fidelity is defined as $F=\text{Tr}[\sqrt{\rho}_{s}\rho\sqrt{\rho}_{s}]$,
the expression for control functions becomes
\begin{align}\label{eq1-6}
  f_{m}(t)=\text{Tr}[\sqrt{\rho_{s}}(-i[H_{m},\rho])\sqrt{\rho_{s}}],
\end{align}
where $\rho_{s}=|S\rangle\langle S|$.

The principle to accelerate the evolution by adding ACC fields can be in fact understood as follows.
The Hamiltonian $H_{0}$ is just used
to guarantee that $|S\rangle$ is a steady state according to Eqs. (\ref{eq0-9}) and (\ref{eq0-10}).
While, by adding the ACC fields, it is easy to
find $(H_{0}+H_{a})|S\rangle\neq 0$ (for $\rho\neq\rho_{s}$ corresponding to $t< t_{f}$) which means
$|S\rangle$ is actually not a steady state when $t<t_{f}$.
For $t\rightarrow t_{f}$, according to Eq. (\ref{eq1-5}), we have $f_{m}(t_{f})=0$ since $\rho|_{t=t_{f}}\rightarrow \rho_{s}$.
Thus, $H_{a}=0$, so that $|S\rangle$ becomes the unique steady state when $t\rightarrow t_{f}$.
That is, when $t<t_{f}$, the coherent fields and dissipation work together to
drive the system to state $|S\rangle$, while when $t\rightarrow t_{f}$,
the ACC fields vanish and the system becomes steady.
It can also be understood as, in the current scheme, $|S\rangle$
is not a steady state until the population of the whole system is totally transferred to it.

By adding a suitable ACC field, such as, $H_{a}=f_{1}(t)H_{1}=f_{1}(t)\mu_{1}|e\rangle_{A}\langle p|+H.c.$,
the fidelity versus time of the speed-up scheme is plotted in Fig. \ref{fig2} (c). Shown in the figure,
in the speed-up scheme, the time required to stabilize the system seems independent to the decay rates.
For an arbitrary cooperativity $C$, an evolution time $700/g$ seems
enough to stabilize the system when a suitable ACC field is applied. To show this in more detail, we plot $F$ versus $C$ in Fig. \ref{figcp} (d).
We can find, for a relatively large cooperativity, i.e., $C=500$, for the scheme in Ref. \cite{Pra95022317}, an evolution time $t_{f}=5000/g$ is still not so enough to stabilize the system,
but an evolution time $t_{f}=700/g$ is enough for the current speed-up scheme.
Take a comparison between Figs. \ref{figcp} (b) and (c),
the time required in the current speed-up scheme is only about 3 times longer than that in a STIRAP scheme,
while, the fidelity of the speed-up scheme can be higher than that of a STIRAP scheme. Therefore,
the current speed-up scheme can be an alternative choice in practice.

\begin{figure}[t]
 \scalebox{0.3}{\includegraphics {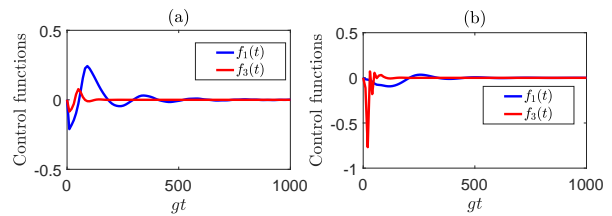}}
 \caption{
         The control functions given according to Eq. (\ref{eq1-5}) for the accelerated dynamics with different initial states.
         The blue-solid curves correspond to the situation that only ACC Hamiltonian $H_{1}$ is added and the red-solid curves
         correspond to the situation that only $H_{3}$ is added.
         (a) The initial state is chosen with $o=1$ and intensity-dependent coefficient is $\mu_{1}=0.3g$.
         (b) The initial state is chosen with $o=0.02$ and intensity-dependent coefficient is $\mu_{3}=0.2g$.
         Parameters are $\Omega=0.07g$, $\omega=0.02g$, $\Xi=5g$, and $\Delta=100g$.
         The decay rates are $\gamma=0.1g$, $\kappa=0$, and $\Gamma=0.001g$.
         }
 \label{fig3a}
\end{figure}

\section{analysis and discussion on the accelerated steady-state entanglement generation}
\begin{figure}[t]
 \scalebox{0.28}{\includegraphics {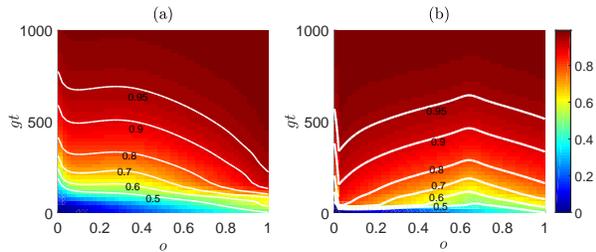}}
 \caption{
         The fidelity of singlet state $|S\rangle$ is plotted as a function of $o$, where $o$ is given
         according to the initial state $\rho_{0}=[o|eg\rangle\langle eg|+(1-o)|gg\rangle\langle gg|]\otimes|0\rangle_{c}\langle 0|$ ($o\in[0,1]$).
         (a) The ACC field is $f_{1}(t)H_{1}$ with $\mu_{1}=0.3g$.
         (b) The ACC field is $f_{3}(t)H_{3}$ with $\mu_{3}=0.2g$.
         parameters are $\Omega=0.07g$, $\omega=0.02g$, $\Xi=5g$, and $\Delta=100g$.
         The decay rates are $\gamma=0.1g$, $\kappa=0$, and $\Gamma=0.001g$.
         }
 \label{fig4}
\end{figure}

First of all, we would like to study how the four ACC Hamiltonians behave in accelerating the entanglement generation.
To ensure that the conditions for obtaining the effective Hamiltonian in Eq. (\ref{eq0-9}) are satisfied,
we choose parameters $\Omega=0.07g$, $\omega=0.02g$, $\Xi=5g$, and $\Delta=100g$.
In the following analysis, the initial state for the system is assumed as
\begin{align}\label{eqrho0}
  \rho_{0}=[o|eg\rangle\langle eg|+(1-o)|gg\rangle\langle gg|]\otimes|0\rangle_{c}\langle0|,
\end{align}
where $o$ is an undetermined coefficient.
We independently display the fidelity of the singlet state $|S\rangle$ versus time in Fig. \ref{fig2} (a) when the
ACC Hamiltonians $H_{1}$ [see the blue-dash curve]
and $H_{3}$ [see the red-dot curve] are added. The effect of $H_{2}$ ($H_{4}$)
is similar with $H_{1}$ ($H_{3}$) on the evolution that does not deserve a separate discussion.
The initial state is chosen as $\rho_{0}=|eg\rangle\langle eg|\otimes|0\rangle_{c}\langle 0|$ in Fig. \ref{fig2} (a).
Shown in the figure, by adding the ACC Hamiltonian $H_{1}$, the entanglement generation is significantly
accelerated ($gt=250$ is enough for a fidelity $\geq 95\%$), while, by adding $H_{3}$, the evolution is almost unchanged.
That is, when the initial state is chosen with $o=1$, the ACC Hamiltonian $H_{3}$ ($H_{4}$)
is unable to accelerate the evolution.
When we change the initial state to
$\rho_{0}=0.02|eg\rangle\langle eg|\otimes|0\rangle_{c}\langle 0|+0.98|eg\rangle\langle eg|\otimes|0\rangle_{c}\langle 0|$
(the following discussion shows that $o=0.02$ is the best choice in this case),
the result becomes different [see Fig. \ref{fig2} (b)] that $H_{3}$ ($H_{4}$) can accelerate the evolution while $H_{1}$ ($H_{2}$) can not.
This result can be understood by Fig. \ref{fig3a} where the corresponding control functions are plotted.
Figures \ref{fig3a} (a) and (b) are plotted with initial conditions $o=1$ and $o=0.02$, respectively.
In Fig. \ref{fig3a}, the blue-solid curves represent the control function $f_{1}(t)$ versus time under different conditions, and the red-solid curves
represent $f_{3}(t)$ versus time. As we can find, the red-solid curve
in Fig. \ref{fig3a} (a) and the blue-solid curve in Fig. \ref{fig3a} (b)
are close to the zero-line, which means the corresponding ACC fields are too weak to accelerate the dynamics.
The blue-solid curve in Fig. \ref{fig3a} (a) and the red-solid curve in Fig. \ref{fig3a} (b)
vanish gradually in an oscillating way along with the increasing of time.
This verifies the ACC fields vanish after a certain evolution time so that
the final stability of the system is guaranteed.
The comparison between Figs. \ref{fig3a} (a) and (b) shows us that $H_{1}$ is
a better choice than $H_{3}$ to be chosen for the accelerated dynamics because the
shape of $f_{1}(t)$ in Fig. \ref{fig3a} (a) is easier to realize than that of $f_{3}(t)$ in Fig. \ref{fig3a} (b).
More can be found from Fig. \ref{fig2} is the choice of ACC Hamiltonian dependents on the initial state.
This point also can be demonstrated by Fig. \ref{fig4} that shows the relationship between the fidelity and the initial state.
According to Fig. \ref{fig4}, for the ACC Hamiltonian $H_{1}$, the evolution is accelerated more remarkably when $o$ is closer to $1$, while
for $H_{3}$, the best choice is $o\rightarrow0.02$.
Besides, the comparison between Figs. \ref{fig2} (a) and (b) [or Figs. \ref{fig4} (a) and (b)] also demonstrates that
the ACC Hamiltonian $H_{1}$ ($H_{2}$) behaves better than $H_{3}$ ($H_{4}$)
in accelerating the evolution.

\begin{figure}[t]
 \scalebox{0.3}{\includegraphics {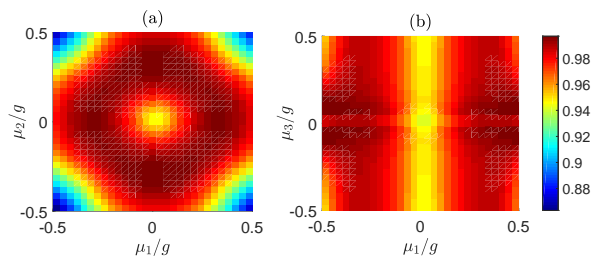}}
 \caption{
         The combined effect of different ACC Hamiltonians on the accelerated dynamics.
         (a) The fidelity of state $|S\rangle$ at the time $t=500/g$ versus ACC fields' intensities $\mu_{1}$ and $\mu_{2}$.
         (b) The fidelity of state $|S\rangle$ at the time $t=500/g$ versus ACC fields' intensities $\mu_{1}$ and $\mu_{3}$.
         parameters are $\Omega=0.07g$, $\omega=0.02g$, $\Xi=5g$, and $\Delta=100g$.
         The decay rates are $\gamma=0.1g$, $\kappa=0$, and $\Gamma=0.001g$.
         }
 \label{fig2a}
\end{figure}

The combined effect of ACC Hamiltonians $H_{1}$ and $H_{2}$ on the accelerated dynamics [see Fig. \ref{fig2a} (a)]
shows the acceleration effect can not be improved by simply adding more same-type ACC
fields or increasing the pulse intensity. The combined effect of
different-type ACC Hamiltonians, i.e., $H_{1}$ and $H_{3}$, is given in Fig. \ref{fig2a} (b).
As compared to Fig. \ref{fig2a} (a), adding different-type ACC Hamiltonians simultaneously has the
ability to slightly improve the fidelity, i.e., $F\simeq 99\%$ when $\mu_{1}\approx0.3g$ and $\mu_{3}\approx 0.1g$.
That is, a high-fidelity steady-state entanglement generation is achievable by suitably choosing ACC fields with suitable intensities.
However, the operation complexity may increase when adding more ACC fields.
So, for convenience, in the following, we focus on analyzing the accelerated entanglement generation by adding the single ACC Hamiltonian $H_{1}$.

\begin{figure}[t]
 \scalebox{0.28}{\includegraphics {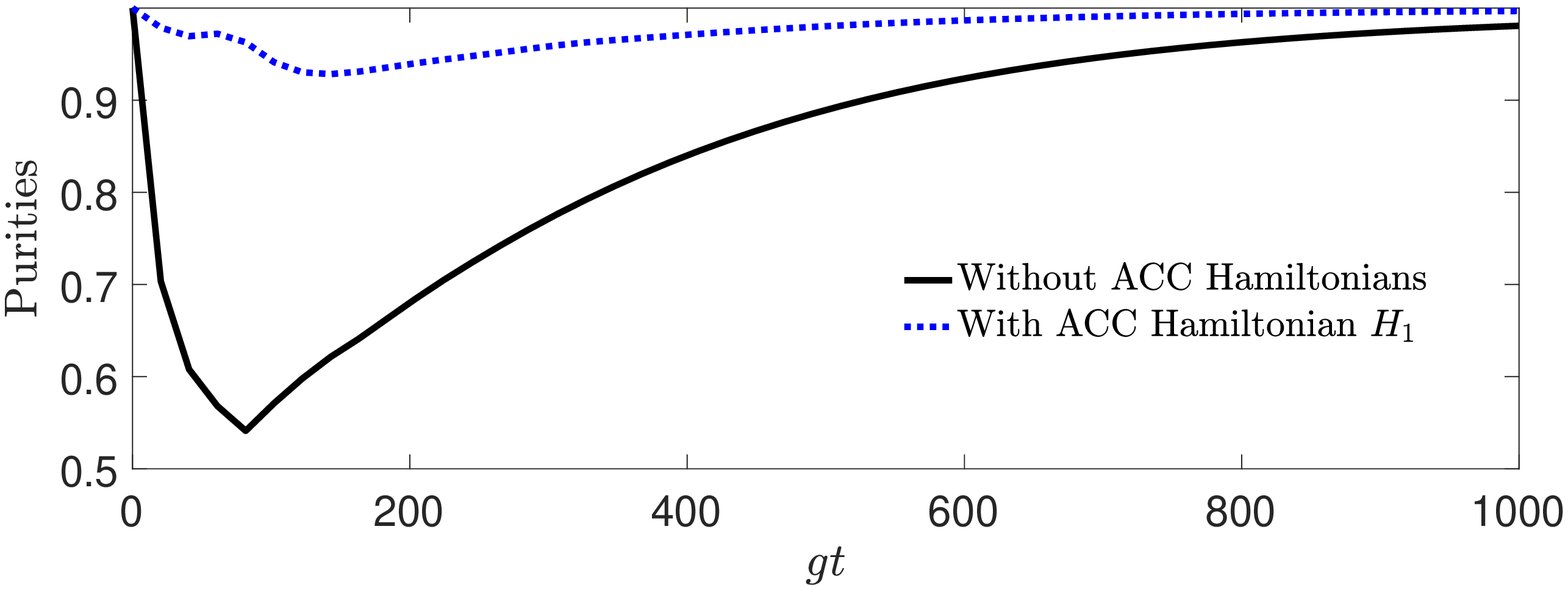}}
 \caption{
         The purity $P(t)=\text{Tr}[\rho^{2}]$ is plotted as a function of time with initial state $|eg\rangle\otimes|0\rangle_{c}$.
         The black-solid and blue-dash curves denote the purities for the general dissipation dynamics and the current accelerated dynamics, respectively.
         Initial state is $|eg\rangle\otimes|{0}\rangle_{c}$ and parameters are parameters are $\Omega=0.07g$, $\omega=0.02g$, $\Xi=5g$, $\Delta=100g$, and $\mu_{1}=0.3g$.
         The decay rates are $\gamma=0.1g$, $\kappa=0$, and $\Gamma=0.001g$.
         }
 \label{fig3}
\end{figure}

\begin{figure}[b]
 \scalebox{0.3}{\includegraphics {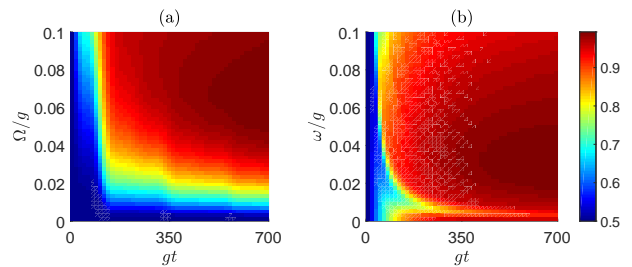}}
 \caption{
         The choice of optimal parameters $\Omega$ and $\omega$ for the accelerated dynamics.
         (a) The fidelity versus $\Omega$ and $gt$.
         (b) The fidelity versus $\omega$ and $gt$.
         Initial state is $|eg\rangle\otimes|{0}\rangle_{c}$ and parameters are $\Xi=5g$, $\Delta=100g$, and $\mu_{1}=0.3g$.
         The decay rates are $\gamma=0.1g$, $\kappa=0$, and $\Gamma=0.001g$.
         }
 \label{fig5}
\end{figure}

We define the purity of a quantum system as $P(t)=\text{Tr}[\rho^{2}]$.
The time evolution of purities for the system with and without the ACC Hamiltonian $H_{1}$
are plotted in Fig. \ref{fig3}. %as blue-dash and black-solid curves, respectively.
We can find from the figure that
the ACC Hamiltonian $H_{1}$ in fact protects the system from dissipation for a certain period of time,
so that the starting point for convergence process is higher than that in a system without ACC Hamiltonians
[the lowest purities in Figs. \ref{fig3} (a) and (b) are about 0.96 and 0.55, respectively].
Hence, the convergence time is shortened.
In Figs. \ref{fig5} (a) and (b), we display the fidelities of state $|S\rangle$ versus $\Omega$ and $\omega$, respectively.
The result shows, the ACC Hamiltonian $H_{1}$ behaves the best in accelerating the entanglement generation when
the Zeno requirement is just satisfied: $\Omega\sim0.1g$ and $\omega\sim0.05g$.
Although the Zeno requirement is fulfilled better with smaller $\Omega$ and $\omega$,
the evolution time is unacceptable long.
The reason can be understood by: when the Rabi frequency $\Omega$ is too small, the system is
slowly excited to the effective excited state $|\phi_{0}\rangle$. As shown in the effective
transitions of the system [see Fig. \ref{fig0} (b)], a certain population for the effective excited state $|\phi_{0}\rangle$ is
necessary for the convergence process, the convergence time will be
long if it is too slow to excite the system to $|\phi_{0}\rangle$.

For the current available parameters
in the cavity QED with Rydberg-blocked atoms \cite{Nat450268,Prl110090402,Njp16043020},
the strength coupling the transition between atomic ground
level $5S_{1/2}$ and the optical level $5P_{3/2}$ of $^{87}$Rb atom to
the quantized cavity mode is $g/2\pi=55$ MHz, the decay
rate of the intermediate state $|p\rangle$ is $\gamma/2\pi=3$ MHz, the
decay rate of the cavity mode is $\kappa/2\pi=1$ MHz, and the
spontaneous emission rate for the Rydberg state $95d_{5/2}$ of
$^{87}$Rb atom is $=0.03$ MHz. By modulating the Rabi
frequencies, detuning parameter and Rydberg interaction
strength satisfying $\Omega=0.07g$, $\omega=0.02g$, and $\lambda=0.5 g$,
the time required to generate a high-fidelity ($\geq 98\%$) steady-state entanglement
is only about $1.5\mu$s ($t_{f}\sim500/g$ and $g=55\times2\pi$ MHz).

\section{Robustness against stochastic parameter fluctuations}
Figure \ref{fig5} in fact indirectly demonstrates that the current accelerated scheme is robust against the systematic errors.
The systematic errors are caused by fixed fluctuations on the parameters.
For example, the fluctuation of Rabi frequency $\Omega$ can be assumed as a fixed value $\delta\Omega=\Omega'-\Omega$
with $\Omega'$ being the real value in experiment.
As shown in Fig. \ref{fig5}, when $\Omega\sim 0.07g$ and $\omega\sim 0.02g$,
the fidelity keeps almost unchanged with the slight changes of $\Omega$ and $\omega$.
That is, the system is robust against systematic errors.
Therefore, in this section, we focus on analyzing the influence of a stochastic kind of noise on the fidelity.
Assume that the Hamiltonian $H_{0}$ is perturbed by
some stochastic part $\eta H_{s}$ describing amplitude noise. A stochastic
Shr\"{o}dinger equation in a close system (in the Stratonovich sense) is then
$\dot{\psi}(t)=[H_{0}+\eta H_{s}\xi(t)]\psi(t)$,
where $\xi(t)=\partial_{t}W_{t}$ is heuristically the time derivative of the Brownian motion $W_{t}$.
$\xi(t)$ satisfies $\langle\xi(t)\rangle=0$ and $\langle\xi(t)\xi(t')\rangle=\delta(t-t')$ because the noise should have zero mean and the noise
at different times should be uncorrelated. Then, we define $\rho_{\xi}(t)=|\psi_{\xi}(t)\rangle\langle\psi_{\xi}(t)|$, and
the dynamical equation without dissipation terms for $\rho_{\xi}$ is thus given as
\begin{align}\label{eq4-1}
  \dot{\rho}_{\xi}=-i[H_{0},\rho_{\xi}]-{i\eta}[H_{s},\xi\rho_{\xi}].
\end{align}
After averaging over the noise, Eq. (\ref{eq4-1}) becomes
\begin{align}\label{eq4-2}
  \dot{\rho}\simeq-i[H_{0},\rho]-{i\eta}[H_{s},\langle\xi\rho_{\xi}\rangle],
\end{align}
where $\rho=\langle\rho_{\xi}\rangle$  \cite{Njp14093040}. According to Novikov's theorem in case of white noise,
we have $\langle\xi\rho_{\xi}\rangle=\frac{1}{2}\langle\frac{\delta\rho_{\xi}}{\delta\xi(t')}\rangle|_{t'=t}=-\frac{i\eta}{2}[H_{s},\rho]$.
Thus, when both the noise and the dissipation are taken into account, the system evolution is governed by
\begin{align}\label{eq4-3}
  \dot{\rho}=&-i[H_{0},\rho]+\mathcal{L}\rho+\mathcal{N}\rho,
\end{align}
where $\mathcal{N}\rho=-\eta^{2}[H_{s},[H_{s},\rho]]/2$. Adding the ACC Hamiltonians, Eq. (\ref{eq4-3}) becomes
\begin{align}\label{eq4-4}
  \dot{\rho}=&-i[H_{0}+H_{a},\rho]+\mathcal{L}\rho+\mathcal{N}\rho.
\end{align}
We choose $H_{a}=f_{1}(t)H_{1}$ in the following analysis. Beware that the control function $f_{1}(t)$ is
given according the master equation in Eq. (\ref{eq1-3}) without the noise terms.

\begin{figure}[t]
 \scalebox{0.28}{\includegraphics {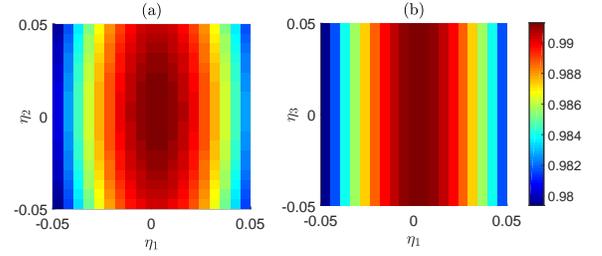}}
 \caption{
         The robustness of the accelerated dynamics against stochastic parameter fluctuations when $t_{f}=700/g$.
         (a) The fidelity versus $\eta_{1}$ and $\eta_{2}$ denoting the amplitude noise intensities of $\Omega$ and $\omega$, respectively.
         (b) The fidelity versus $\eta_{1}$ and $\eta_{3}$ denoting the amplitude noise intensities of $\Omega$ and $U_{rr}$, respectively.
         Initial state is $|eg\rangle\otimes|{0}\rangle_{c}$ and parameters are $\Omega=0.07g$, $\omega=0.02g$, $\Xi=5g$, $\Delta=100g$, and $\mu_{1}=0.3g$.
         The decay rates are $\gamma=0.1g$, $\kappa=0$, and $\Gamma=0.001g$.
         }
 \label{fig6}
\end{figure}

For the current scheme, we consider the amplitude noises exist in
\begin{align}\label{eq4-5}
  H_{s1}=&\Omega|p\rangle_{A}\langle e|+H.c., \cr
  H_{s2}=&\omega|g\rangle_{A}\langle e|+H.c., \cr
  H_{s3}=&U_{rr}|rr\rangle\langle rr|,
\end{align}
with intensities $\eta_{1}^{2}$, $\eta_{2}^{2}$, and $\eta_{3}^{2}$, respectively.
The last line in Eq. (\ref{eq4-5}) is considered because it is difficult to
accurately adjust the distance between two Rydberg atoms in experiment.
In Fig. \ref{fig6}, we simulate the steady state
fidelity as a function of $\eta_{1,(2,3)}$ to analyze the influence of amplitude noises.
Fortunately, the current scheme is robust against the amplitude noises caused by the microwave field and the Rydberg-mediated interaction,
the scheme permits $\eta_{2,(3)}\in[-5\%,5\%]$ so as to preserve the
fidelity almost unchanged. For error Hamiltonian $H_{s1}$, the negative effect of amplitude noise on the fidelity is also very small,
only $1\%$ deviation is caused even when the noise intensities are $\eta_{1}=\eta_{2}=\eta_{3}=0.1$.
That is, the accelerated scheme is demonstrated to be robustness against amplitude-noise errors.

\section{conclusion}
In conclusion, we have proposed a scheme based on Lyapunov control to accelerate the generation
of steady-state entanglement in a cavity QED system with Rydberg atoms.
The ACC fields in fact protect the system from dissipation in a certain time. Thus an imperfect unitary evolution is allowed
for the system to rapidly reach the target steady state with fidelity about $90\%$ in the first evolution stage. Then, in the second evolution stage, the
ACC fields gradually vanish and the dissipation dynamics occupies a leading
position to converge the system to target steady state (from fidelity $\sim 90\%$ to $\sim 100\%$).
Numerical simulation demonstrates that the time required for entanglement generation with fidelity $\geq 95\%$ has been
shortened by about 6 times as compared to that for a scheme without ACC fields.
Moreover, the accelerated scheme is robust against noise errors as demonstrated by numerical simulation.
As a result, the current scheme combining the advantages of coherent unitary dynamics and dissipation dynamics
allows for significant improvement in quantum entanglement generation.
Therefore, we hope that the current work may open venues for the
experimental realization of entanglement in the near future.

\section*{Acknowledgements}
This work was supported by the National Natural Science Foundation
of China under Grants No. 11575045, No. 11374054 and No. 11675046.

\section*{Appendix. Quantum Zeno dynamics}
The quantum Zeno effect which has
been tested in many experiments is the inhibition of transitions
between quantum states by frequent measurements \cite{Jmp18756,Pra412295,Prl744763,Pst2149}. It shows that a
system can actually evolve away from its initial state
while it still remains in the so-called Zeno subspace determined
by the measurement when frequently projected
onto a multidimensional subspace. This was called ``quantum
Zeno dynamics'' (QZD) by Facchi and Pascazio in 2002 \cite{Prl89080401Jpa41493001}.
In fact, QZD can be achieved via
continuous coupling between the system and an external
system instead of discontinuous measurements.
Here, we give an elementary introduction to this kind of QZD.
A generic Hamiltonian of a dynamical evolution can be written as
\begin{align}\label{eq-A1}
  H=H_{c}+KH_{p},
  \tag{A1}
\end{align}
where $H_{c}$ is the Hamiltonian of the quantum system, $H_{p}$ is
an interaction Hamiltonian caricaturing the continuous
measurement, and $K$ is coupling constant. In the strong coupling limit, $K\rightarrow \infty$,
the subsystem of interest is dominated by the evolution operator
\begin{align}\label{eq-A2}
  U_{0}(t)&=\lim_{K\rightarrow\infty}\exp{(i K H_{p}t)}U(t)\cr
          &=\exp{(-it\sum_{n}P_{n}H_{c}P_{n})},
  \tag{A2}
\end{align}
where $P_{n}$ is the projector onto the space of eigenstates of $H_{p}$ with
eigenvalues $\zeta_{n}$, i.e., $H_{p}=\sum_{n}\zeta_{n}P_{n}$. Thus, the whole system is governed
by the limiting evolution operator
\begin{align}\label{eq-A3}
  U(t)=&\exp(-iK H_{p}t)U_{0}(t)\cr
      =&\exp[-it\sum_{n}(K\zeta_{n}P_{n}+P_{n}H_{c}P_{n})].
      \tag{A3}
\end{align}
The effective Hamiltonian (also known as the ``Zeno Hamiltonian'') for the system is accordingly given as
\begin{align}\label{eq-A4}
  H_{Z}=\sum_{n}(K\zeta_{n}P_{n}+P_{n}H_{c}P_{n}).
  \tag{A4}
\end{align}
In the current scheme, we consider $H_{ac}$ as $KH_{p}$ and $H_{r}$ as $H_{c}$,
the strong coupling limit $K\rightarrow\infty$ corresponds to $g\gg \Omega$. According to Eq. (\ref{eq-A4}) and Ref. \cite{Pra95022317}, the
effective Hamiltonian in Eq. (\ref{eq0-9}) can be obtained.

\end{document}